\documentclass[nohyperref]{article}

\usepackage{microtype}
\usepackage{graphicx}
\usepackage{subfigure}
\usepackage{booktabs} 

\usepackage{hyperref}

\usepackage[accepted]{icml2022}

\usepackage{amsmath}
\usepackage{amssymb}
\usepackage{mathtools}
\usepackage{amsthm}

\usepackage[capitalize,noabbrev]{cleveref}

\theoremstyle{plain}

\theoremstyle{definition}

\theoremstyle{remark}

\usepackage[textsize=tiny]{todonotes}

\icmltitlerunning{Ad Hoc Teamwork in the Presence of Adversaries}

\begin{document}

\twocolumn[
\icmltitle{Ad Hoc Teamwork in the Presence of Adversaries}




\begin{icmlauthorlist}
\icmlauthor{Ted Fujimoto}{pnnl}
\icmlauthor{Samrat Chatterjee}{pnnl}
\icmlauthor{Auroop Ganguly}{ne,pnnl}


\end{icmlauthorlist}

\icmlaffiliation{pnnl}{Pacific Northwest National Laboratory, WA, USA}
\icmlaffiliation{ne}{Northeastern University, MA, USA}

\icmlcorrespondingauthor{Ted Fujimoto}{ted.fujimoto@pnnl.gov}

\icmlkeywords{Machine Learning, ICML}

\vskip 0.3in
]



\printAffiliationsAndNotice{}  

\begin{abstract}
Advances in ad hoc teamwork have the potential to create agents that collaborate robustly in real-world applications. Agents deployed in the real world, however, are vulnerable to adversaries with the intent to subvert them. There has been little research in ad hoc teamwork that assumes the presence of adversaries. We explain the importance of extending ad hoc teamwork to include the presence of adversaries and clarify why this problem is difficult. We then propose some directions for new research opportunities in ad hoc teamwork that leads to more robust multi-agent cyber-physical infrastructure systems. 
\end{abstract}

\section{Introduction}

Ad hoc teamwork (AHT) is a step towards general multi-agent intelligence where decision-making agents (human and computational) are trained to collaborate with other previously unseen agents. That is, AHT assumes that each agent will learn to collaborate without prior coordination or direct control of the other agents \cite{mirsky2022survey}. Although the challenge of AHT was first formalized by \citet{stone2010ad}, there has been a recent resurgence of AHT research that incorporates advances in deep reinforcement learning (RL).

Established works have focused on multi-agent consensus among a network of agents typically with centralized training and decentralized execution, where a fixed set of agents are known to all \cite{lowe2017multi, rashid2018qmix}. For example, this may be useful in real-world infrastructure protection settings where a team of collaborative defense agents might be tasked with securing an asset of interest. This might not be practical, however, in dynamic, adversarial, and contested environments. In these settings, new agents may need to enter the environment and start competing or collaborating without access to a centralized policy. Even if we assume such a policy, it would likely lead to brittle agents that are tethered to their training partners for optimal performance \cite{li2019robust}. This leads the group to be unable to perform optimally in the presence of adversaries. One appeal of AHT is the potential for multiple people or organizations to train agents that collaborate with each other without the need to retrain with previously unseen agents. Outside of \citet{shafipour2021task}, AHT in the presence of adversaries has not been explored due to the assumption of full cooperation between agents. Adversarial machine learning (ML) researchers have an opportunity to chart the next path of truly robust multi-agent learning. In this paper, we contribute (1) a summary of past research in AHT, (2) why it could be a necessary next step in high-stakes, real-world multi-agent RL, (3) a list of some challenges in AHT in the presence of adversaries, and (4) a proposal for some new directions for researchers in both adversarial ML and AHT.

\section{Ad Hoc Teamwork}

\subsection{Background}

 AHT is a subfield of cooperative AI, which is defined as AI methods that help individuals, humans and machines, to find ways that improve joint welfare \cite{dafoe2020open}. As mentioned previously, AHT methods train agents to collaborate without prior coordination or direct control of other agents. In this paper, AHT will mostly be synonymous with RL-based methods in AHT, called zero-shot coordination (ZSC). The main reason is because ZSC has enjoyed some recent success. These successes include a general and differentiable information theoretic objective for training a diverse population of optimal policies \cite{lupu2021trajectory}, training methods that prevent agents from exchanging information through arbitrary conventions \cite{hu2021off}, and graph neural networks that learn agent models and joint-action value models under varying team compositions \cite{rahman2021towards}. Some approaches even outperform self-play and behavioral cloning when collaborating with humans in video games \cite{strouse2021collaborating}. ZSC is an evolving field and there is currently some disagreement over how to precisely formulate it \cite{treutlein2021new}. We recommend \citet{mirsky2022survey} as an up-to-date survey of AHT.

\subsection{Potential Applications}

While AHT has shown some success in robot soccer \cite{genter2017three}, there are still many potential applications yet to be explored. Ideally, AHT can perform more robustly than centralized multi-agent RL methods, so any multi-agent environment can potentially train an AHT application. In the context of secure cyber-physical infrastructure operations (such as transportation and energy distribution), multi-agent learning and control is essential for achieving system resilience goals \cite{januario2019distributed, hou2021reinforcement, phan2021resilient}. Some future AHT applications worth mentioning include important automated infrastructure, like energy and power distribution \cite{biagioni2021powergridworld, pigott2021gridlearn, wang2021multi}. AHT would be particularly useful here because of the possibility that different buildings or power plants might need to coordinate with previously unseen agents from other allied power plants. Another application would be human-automated vehicle interaction. It is still difficult for automated cars to learn and appropriately react to humans while driving \cite{dommes2021young}.

\section{Potential Challenges with Adversaries in Ad Hoc Teamwork}

We now briefly describe some challenges that agents trained using AHT methods could face unless it is framed as an adversarial ML problem.

\subsection{Past Research in Adversarial RL}

It is possible that some attacks described in past research on adversarial RL \cite{huang2017adversarial, gleave2019adversarial, lin2020robustness, sun2020stealthy, figura2021adversarial, fujimoto2021reward} will have similar negative effects in AHT. Given that there are currently no universally effective defenses against adversarial attacks in supervised learning \cite{athalye2018obfuscated, short2019defending}, the same is likely true for RL.

\subsection{Balancing the Presence of Allies and Adversaries}

Even if we assume adversaries might exist in the environment, it is not clear how to train the agents to collaborate. It might be that the adversary is secretly controlling the policy of one of the agents. We would need new methods that assume some of the training agent's partners might not be aligned with the objectives of the group.

\subsection{Methods of Evaluation}

One aspect of AHT that might be overlooked is the evaluation of the agents at test time. An AHT agent might have to act differently in the presence of an adversary instead of a low-skill partner. In general, we would need AHT methods that are more risk-aware and define what success looks like when there is a high probability that some of the other agents will try to subvert the group's goal.

\section{Future Research Directions}

AHT was established with robust multi-agent behavior as the main goal. Some past works in AHT have made steps in that direction. Type-based methods, like in \citet{rahman2021towards}, attempt to infer the agent's behavior from its actions. Agents can also use communication to help each other quickly adapt to new scenarios \cite{macke2021expected}, which as been successful in defending against attacks in multi-agent deep learning \cite{tu2021adversarial}. Extending past work in AHT can show us which methods have the most potential in being robust against adversaries. Past works in adversarial training in RL \cite{pinto2017robust, pan2019risk} show that robustness can be achieved if the agent is trained with adversarial examples.

Extending past methods alone, however, will likely not be enough. Introducing new agents to other previously trained agents exacerbate the brittleness of current RL methods \cite{gleave2019adversarial}. The success of adversarial AHT, and AHT in general, will depend on a foundation of risk-awareness for multi-agent RL in the presence of adversaries. This foundation needs to account for the ``openness" of the environment, which models how new agents enter the group. The presence of adversaries compels us to carefully understand how openness impacts the training of an AHT agent. Openness that includes adversarial agents will require researchers to assume additional formal descriptions of the environment, which include (1) the distribution of new adversarial agents to introduce to the group, (2) the diversity of the adversaries, and (3) the reward function that accounts for the number and types of adversaries. As multi-agent RL is starting to be applied to more real-world tasks \cite{wang2021multi, bae2022scientific}, there is now more motivation to train agents that are robust and risk-aware. This creates an incentive to train AHT agents in high-stakes environments. An adversarial ML perspective can make these endeavors more successful.

\section{Conclusion}

As RL continues to be pursued to accomplish complex, real-world tasks, we will inevitably need to train RL agents that robustly cooperate with others. Since AHT seems to be a promising path toward that direction, we propose accounting for the possible presence of adversaries as a necessary component of that field. We put forward that adversarial ML researchers will be critical to ensuring AHT will advance safe collaboration between both humans and RL agents.


\bibliography{aht_icml2022}
\bibliographystyle{icml2022}

\end{document}